\documentstyle{article}
\newtheorem{D}{Definition}
\newtheorem{T}{Theorem}

\newenvironment{ST}{\paragraph{Stipulation}\em }{\/\rm \par\vskip 
8pt}
\newlength{\picw}
\newlength{\picfr}

\newenvironment{Proof}{\begin{quote}\footnotesize
{\bf Proof}\hskip 0.3cm}{\hbox{ }\hfill $\Box$\end{quote}\normalsize 
}

\begin{document}
\input epsf

\title{\bf IS THERE ANYTHING NON-CLASSICAL?}
\author{L\'ASZL\'O E. SZAB\'O\thanks{E-mail: 
leszabo@ludens.elte.hu} \\{\normalsize {\em Institute for 
Theoretical Physics\/}} \\{\normalsize {\em E\"otv\"os University of 
Budapest\/}} 
}

\date{15 July 1996}
\maketitle
\begin{abstract}
It is argued that quantum logic and quantum probability theory are 
fascinating mathematical theories but without any relevance to our real 
world. 
\end{abstract}

\section{Introduction}

At the beginning of this century, which is sometimes called the century 
of physics,  we were presented with two revolutionary physical 
theories: the theory of relativity and the quantum theory.  The common 
feature of these theories (and, perhaps, of the whole physics in this 
century) is their {\em counterintuitive\/} character. It would be beyond 
the scope of this paper to analyze the intellectual background at the 
turn of the century that made scientists so much attracted by everything 
against intuition. No doubt, in the twenties century physics ``the logical 
basis is getting farther and farther from the empirical data, and the 
mental way leading to theorems directly related to the empirical 
observations is getting more and more long and hard" (Einstein, 1938). 
I believe, however,  that  in quantum mechanics we swung to the other 
extreme by enforcing counterintuitive abstractions, instead of 
struggling for real explanations. 

The target of my critique is the {\em quantum probability theory\/} and 
{\em quantum logic\/}. The story begins with von Neumann's 
recognition\footnote{This idea first appeared in Neumann 1932. One 
can find it in a more explicit and somewhat different form in Birkhoff 
and Neumann 1936.}, that
quantum mechanics {\em can be regarded\/} as a kind of  probability 
theory defined over the subspace lattice $L(H)$  of a Hilbert space 
$H$.
This recognition was confirmed by the {\em Gleason theorem\/}: 
\begin{D}
\label{probmeasure}
A non negative real function $\mu$ on $L(H)$ is called a probability 
measure if $\mu (H)=1$ and if whenever $E_{1},E_{2}, \ldots $ are 
pairwise orthogonal subspaces, and $E=\bigvee^{\infty }_{i=1}E_{i}$, 
then   $\mu (E)=\sum^{\infty }_{i=1}\mu(E_{i})$.
\end{D}

\begin{T}[Gleason 1957]
\label{Gleason}
If  $ H $ is a real or complex Hilbert space of dimension greater than 2, 
and $\mu$ is a probability measure on $L(H)$, then there exists a 
density operator  $ W $ on  $ H $, such that $(\forall E\in L(H))\left[ 
\mu(E)=tr(WE)\right] $.\footnote{The subspaces, the corresponding 
projectors and the corresponding events are denoted by the same 
letter.}
\end{T}

Formally, the intersection   and the (closed) linear union  of subspaces 
play the role of {\em conjunction\/} and {\em disjunction\/} in an 
underlying {\em event\/} lattice of a {\em probability theory\/}. In spite 
of some difficulties, it is commonly accepted that, for example, a 
conjunction $A\wedge  B $, represented by the intersection of the 
corresponding subspaces, $A \cap B$,   corresponds to an event, which 
is nothing else, but the {\em joint occurrence\/} of events $ A $ and $ 
B $. However,  we have, as we will see it soon, many difficulties with 
such an interpretation! While everything is clear from the mathematical 
point of view,  the {\em physical\/} meaning assigned to the elements 
of the subspace lattice and to the lattice operations is far from obvious.

The quantum probability --- quantum logic approach is based on the 
conviction that there are phenomena described by quantum mechanics 
which cannot be accommodated in the classical Kolmogorov theory of 
probability. The majority of authors are shearing this conviction: Jauch 
(1968), Bub (1974), Putnam (1974), Piron (1976), Mittelstaedt (1978), 
Beltrametti and Cassinelli (1981), Gudder (1988), Pitowsky (1989), 
K\"ummerer and Maassen (1996) and many others. Due to Feynmann 
(1951) this opinion is quite common among the users of quantum 
mechanics but who do not care to much with foundational questions.

There has been a serious critique against this approach. The first paper 
pointing out the pitfalls of quantum logic was published by Strauss 
(1937) a year after the famous Birkhoff and Neumann (1936). It is also 
worth mentioning a few papers of the last years arguing against 
non-classical probabilities: Ballentine (1989), Costantini (1992), 
Szab\'o (1995a,b), Gill (1996). 
It seems to me that quantum probabilitists completely ignore the 
serious pitfalls pointed out by these authors. Bell expressed quite a 
similar disappointment: 
\begin{quote}
{\it Why did such serious people take so
seriously axioms which now seem so arbitrary? I suspect that they were
misled by the pernicious misuse of the word `measurement' in 
contemporary
theory. This word very strongly suggests the ascertaining of some
preexisting property of some thing, any instrument involved playing a
purely passive role. Quantum experiments are just not like that, as we
learned especially from Bohr. The results have to be regarded as the 
joint product of `system' and `apparatus,' the complete experimental 
set-up. But the misuse of the word `measurement' makes it easy to 
forget this and then to expect that the `results of measurements' should 
obey some simple logic in which the apparatus is not mentioned. The 
resulting difficulties soon show that any such logic is not ordinary 
logic.  It is my impression that the whole vast subject of `Quantum 
Logic' has arisen in this way from the misuse of a word. I am 
convinced that the word `measurement' has now been
so abused that the field would be significantly advanced by banning its 
use altogether, in favor for example of the word `experiment.'}\hfill  
 (Bell 1987, p. 166.)
\end{quote}

My aim is to show in this paper that, beyond their vagueness and 
counterintuitiveness, quantum logic and quantum probability theory are 
{\em needless and inadequate\/}, because there is nothing in reality 
described by these mathematical constructions.  In 
Section~\ref{nonsensical} I will show that whenever we consider 
non-commuting element of $L(H)$ nonsensical ``probabilities" can 
appear which can hardly be interpreted as relative frequencies of any 
events. In Section~\ref{need} two often quoted examples, the double 
slit experiment and the EPR experiment are analyzed. These examples 
are usually meant to illustrate why we need to create a new, 
non-classical theory of probability. In both cases, however, it will be 
shown that no need to supersede the Kolmogorov theory of probability. 
In Section~\ref{Howtojoin} we will see how can quantum phenomena, 
in general, be accommodated in the classical Kolmogorov theory of 
probability.

\section{Nonsensical probabilities}
\label{nonsensical}

Sometimes quantum mechanics produces very strange values of 
``probabilities". Let me give a simple example. 

\subsection{Example I}
Consider a system described in a 2-dimensional Hilbert space 
$H^{2}$. Let $\varphi $ and $\psi $ be two orthogonal unit vectors. 
Suppose that the system is in the pure state $W=P_{\varphi }$. 
Consider non-commuting elements of $L(H^{2})$: let $E_{1}$ be the 
one-dimensional subspace spanned by $(cos\,\theta )\varphi + 
(sin\,\theta )\psi $ and $E_{2}$ the subspace spanned by $(cos\,\theta 
)\varphi -(sin\,\theta )\psi $. The intersection is $E_{1}\cap 
E_{2}=\emptyset $. The probabilities of the corresponding events are 
\begin{eqnarray*}
p(E_{1})&=&tr(WE_{1})=cos^{2}\,\theta , \\
p(E_{2})&=&tr(WE_{2})=cos^{2}\,\theta ,\\
p(E_{1}\wedge E_{2})&=&tr(W(E_{1}\cap E_{2}))=0.
\end{eqnarray*}

If $\theta $ is close to zero then, for example,
$$ p(E_{1})=0.9,\: p(E_{2})=0.9,\: {\rm while} \: p(E_{1}\wedge 
E_{2})=0.$$

It is, however, {\em impossible\/} to interpret similar numbers as {\em 
relative  frequencies\/} of occurrences of $E_{1},\,E_{2}$ and 
$E_{1}\wedge E_{2}$. One cannot classify the possible histories of 
the universe in such a way, that  90\% of them contains event $E_{1}$,  
90\% of them contains $E_{2}$, but none of them contains both 
$E_{1}$ and $E_{2}$.

It is easy to see, that probabilities cannot be interpreted as relative 
frequencies if the following inequality is violated:
\begin{eqnarray}
\label{1}
p(E_{1})+p(E_{2})-p(E_{1}\wedge E_{2})\leq 1.
\end{eqnarray}

To better understand the significance of the above example it is 
necessary to make a few remarks:
\begin{itemize}
\item If $E_{1}$ and $E_{2}$ commute then inequality (\ref{1}) is 
always satisfied.

\item At first sight one could think that the difficulties related with 
conjunctions occur only if the system is  in a special state. But, it is 
easy to see that {\em for each state\/} one can find elements of 
$L(H)$,  for which inequality (\ref{1})  is violated.

\item From the point of view of violation  of inequality (\ref{1}) the  
{\em complementarity\/} of $E_{1}$ and  $E_{2}$ is
irrelevant. One can easily create a similar example in space 
$H^{3}$), such  that $ E_{1}\wedge 
E_{2}\neq \emptyset $ but inequality (\ref{1} is violated.  The 
violation  of inequality (\ref{1})  means a more serious difficulty than 
the problem of complementary observables. 
 
\item Equation 
\begin{eqnarray}
\label{neumann}
p(E_{1}\vee E_{2})= p(E_{1})+ p(E_{2})- p(E_{1}\wedge 
E_{2})
\end{eqnarray}
implies inequality (\ref{1}).  Remarkably, von Neumann 
regarded (\ref{neumann}) as a  fundamental property of a 
probability measure, which should be required in the quantum 
case, too.(Cf. R\'edei, forthcoming)  
\end{itemize}

\subsection{Does non-commutativity always bear the danger of 
nonsensical probabilities?}

It is not surprising that the appearance of nonsensical probabilities is  
related with non-commutative projectors. I must, however, emphasize 
that the naturality of this fact does not justify at all the adherence to 
nonsensical probabilities. 
\begin{T}
\label{en}
Let $E_{1}$ and $E_{2}$ be non-commuting elements of $L(H)$. 
There  exists a pure state $\Psi $ for which the probabilities violate 
inequality (\ref{1}):
$$\left\langle \Psi ,E_{1}\Psi \right\rangle + \left\langle \Psi ,E_{2}\Psi 
\right\rangle - \left\langle \Psi ,(E_{1}\wedge E_{2}\Psi) \right\rangle 
>1.$$
\end{T}
\begin{Proof}
\normalsize

Arbitrary  $ E_{1}$  and $ E_{2}$  can be written in the following 
form:
\begin{eqnarray}
\label{biz1}
E_{1}=( E_{1}\wedge E_{2})\vee A, \nonumber\\
E_{2}=( E_{1}\wedge E_{2})\vee B,
\end{eqnarray}
such that $A\bot E_{1}\wedge E_{2}$ and $B\bot E_{1}\wedge 
E_{2}$.
		
First we prove the following statements:
\begin{itemize}
\item[(a)] $A\neq \emptyset $ and $ B\neq \emptyset $ and $A\neq B$.
\item[(b)] $A\not\perp B$.
\end{itemize}

Indeed, if $A= \emptyset $ or $ B= \emptyset $ or $A=B$ would hold 
then either $ E_{1}< E_{2}$ or $ E_{2}< E_{1}$, that would 
contradict to the assumed non-commutativity of $ E_{1}$ and $ 
E_{2}$. For proving (b) we show that from $A\bot B$  also the 
commutativity of  $ E_{1}$ and $ E_{2}$ would follow. 
Commutativity is equivalent with $E_{1}=( E_{1}\wedge E_{2})\vee ( 
E_{1}\wedge E_{2}^{\bot })$. Using (\ref{biz1}) we have
\begin{eqnarray}
\label{biz2}
(E_{1}\wedge E_{2})\vee \left[ \left( (E_{1}\wedge E_{2})\vee 
A\right)\wedge \left((E_{1}\wedge E_{2})\vee B \right)^{\bot } \right]   
\nonumber \\
=(E_{1}\wedge E_{2})\vee \left[ \left( (E_{1}\wedge E_{2})\vee 
A\right)\wedge \left((E_{1}\wedge E_{2})^{\bot }\wedge  B^{\bot } 
\right) \right]\\
=(E_{1}\wedge E_{2})\vee \left[ \left( (E_{1}\wedge E_{2})\vee 
A\right)\wedge (E_{1}\wedge E_{2})^{\bot }\right]\wedge  B^{\bot }  
\nonumber 
\end{eqnarray}

Since $A\bot (E_{1}\wedge E_{2})$, the distributivity holds in the 
square brackets. Therefore we can continue (\ref{biz2}) as follows:
\begin{eqnarray*}
=(E_{1}\wedge E_{2})\vee (A\wedge B^{\bot })\\
=(E_{1}\wedge E_{2})\vee A= E_{1}
\end{eqnarray*}
which proves (b).

Now, from (a) and (b) it follows that there exists at least one 
normalized vector $\Psi \in A$ such that $\Psi \not\in B$. Such a $\Psi 
$ is a state vector for which the inequality 
\begin{eqnarray}
\label{durva}
\underbrace{\left\langle \Psi ,E_{1}\Psi \right\rangle }_{1}+ 
\underbrace{\left\langle \Psi ,E_{2}\Psi \right\rangle}_{ >0}- 
\underbrace{\left\langle \Psi ,(E_{1}\wedge E_{2}\Psi) \right\rangle 
}_{0}>1
\end{eqnarray}
 holds. 
\end{Proof}

The strange meaning of (\ref{durva}) is obvious! If $E_{1}$ happens 
with certainty, how can $E_{2}$ occur   without  $E_{1}$? 	

 Thus, our partial conclusion can be  this: 1)~$L(H)$ can hardly play 
the role of  an ``algebra of events" for a probability theory. 2)~The 
number $tr(WE)$ cannot be interpreted as  the  ``relative 
frequency" of an event.

\section{Do we really need quantum probability theory and 
quantum logic?}
\label{need}

\subsection{Example II: The double slit experiment}
Our next example is  the double slit experiment which is often quoted 
in  order to justify why we need  quantum probability theory 
(Fig.~\ref{doubleslitfig}).

\setlength{\picw}{10cm}
\setlength{\picfr}{\picw}
\addtolength{\picfr}{0.5cm}
\begin{figure}
\begin{center}
\makebox[\picfr][c]{\vbox{\centerline{\epsfxsize=\picw 
\epsfbox{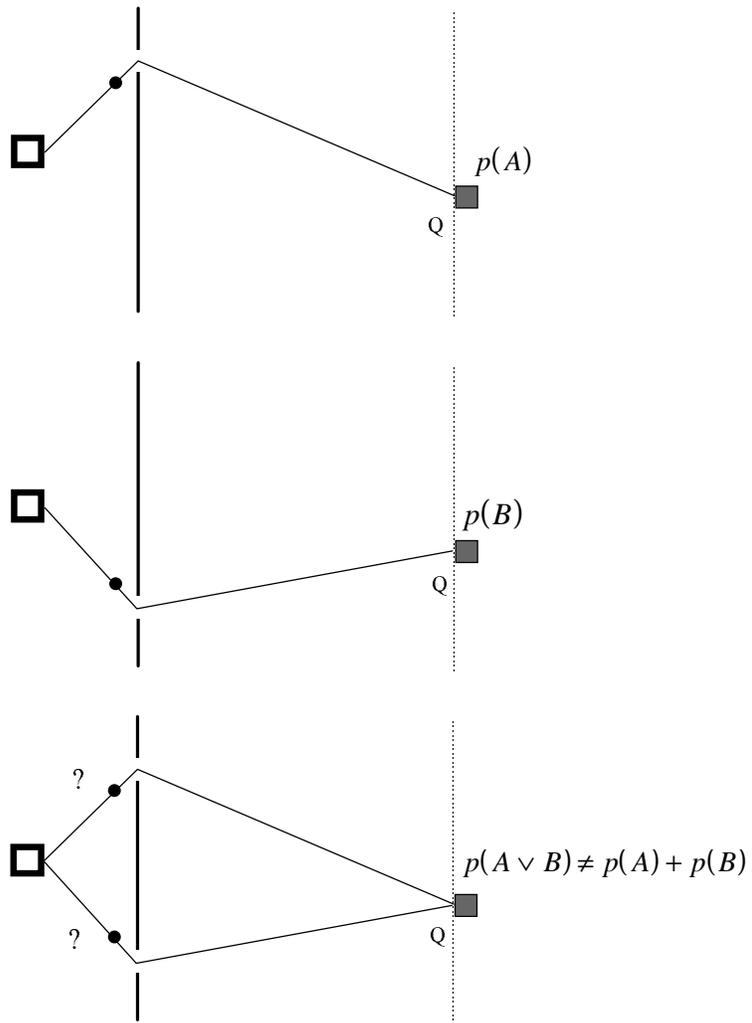}}}}
\end{center}
\caption{According to the usual interpretation the double slit 
experiment indicates that the rules of computing probabilities in 
quantum mechanics must be different from that of classical probability 
theory.}
\label{doubleslitfig}
\end{figure}

Denote $p(A)$ the probability of that ``the particle arrives at a given 
point of the screen, Q, when only slit~1 is open". $p(B)$ denotes the 
similar probability for slit~2. In the experiment one finds that 
\begin{eqnarray}
\label{doulenoneq}
p(A) + p(B) \neq p(A \vee B),
\end{eqnarray}
where $ p(A \vee B)$ stands for the probability of  ``the particle arrives 
at point Q either through slit~1 or slit~2". 
According to the usual interpretation the double slit experiment shows 
that 
``{\it the method of computing probabilities involving subatomic 
particles is different from that of classical probability theory\/}"
(Gudder 1988, p. 57). Therefore we must, as the usual conclusion says, 
1)~change probabilities for complex amplitudes (Feynmann) or 2)~give 
up the Booleanan event lattice (Quantum Logic) and classical 
probability theory(Quantum Probability Theory).

{\em Contrary\/} to these conclusions, let me ask:

\subsection{Why don't we analyze the double slit example more 
carefully? }
There are two different ways in which we can correctly describe the 
double slit experiment within the framework of classical probability 
theory. 
In both cases, it is the precise usage of notions ``event" and 
``disjunction" what makes the classical probability theory satisfactory, 
while the formula (\ref{doulenoneq}) is, as we will see it soon, based 
on the misuse of these notions. 

\subsubsection*{Version I}

We must precisely distinguish the following events:
\begin{itemize}
\item[$A$:] ``Slit 1 is open and slit 2 is closed and the particle is 
detected at $Q$"
\item[$B$:] ``Slit 1 is closed and slit 2 is open and the particle is 
detected at $Q$"
\item[$C$:] ``Slit 1 is open and slit 2 is open and  the particle is 
detected at $Q$"
\end{itemize}
Obviously,
$$A\vee B\neq C.$$
Consequently we are not surprised that 
$$p(A\vee B)=p(A)+p(B)\neq p(C).$$
That is, formula (\ref{doulenoneq}) is incorrect, consequently there is 
no violation of classical rules of probability calculation.

\subsubsection*{Version II}

There is only one event:
\begin{itemize}
\item[$D$:] ``The particle is detected at $Q$"
\end{itemize}
\noindent
There are, however, different {\em conditions\/} under which the 
probabilities are understood.
But the Kolmogorov axioms are meant to apply to probabilities 
belonging to {\em one common\/} system of conditions! 
Consequently, it does not mean a violation of  the Kolmogorov axioms 
if 
\begin{eqnarray*}
p_{ 1\,is\,open;\  2\, is\, closed}(D)+ p_{1\,is\,closed;\  2\,is\,open}(D)
\neq  p_{1\, is\,open;\  2\,is\,open}(D).
\end{eqnarray*}

\subsection{Example III: The EPR experiment}
We have seen in the previous subsection that the double slit experiment 
does not prove  the nonapplicability of Kolmogorov's classical theory 
of probability. It is true, however, that this example is not regarded as a 
serious one: it is rather used in the quantum mechanics text books only. 
In this subsection we are going to analyze the Einstein-Podolsky-Rosen 
experiment which is regarded as a crucial -- empirically tested -- 
situation providing probabilities which do not conform with the 
Kolmogorovian theory. 

\setlength{\picw}{8cm}
\setlength{\picfr}{\picw}
\addtolength{\picfr}{0.5cm}
\begin{figure}
\begin{center}
\makebox[\picfr][c]{\vbox{\centerline{\epsfxsize=\picw 
\epsfbox{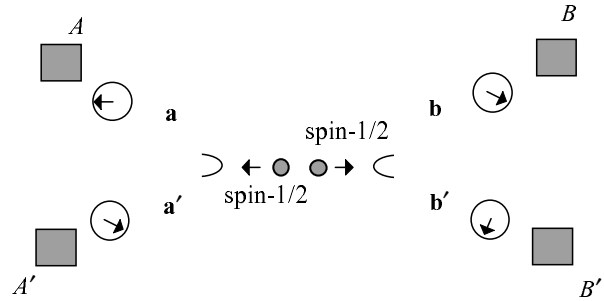}}}}
\end{center}
\caption{The Aspect experiment with spin-$\frac{1}{2}$ particles}
\label{Aspect}
\end{figure}

Consider an Aspect-type EPR experiment
with spin-$\frac{1}{2}$ particles (Fig.~\ref{Aspect}). 
The four detectors detect the spin-up events. The two switches are 
making choice from sending the particles to the Stern-Gerlach magnets 
directed into different directions. The {\em observed\/} events are the 
followings:
\begin{center} 
\begin{tabular}{rl} 
\leavevmode
$A$ :&The ``left particle has spin `up' along direction ${\bf a} $" 
detector 
beeps\\
$A'$ :&The ``left particle has spin `up' along direction ${\bf a'} $" 
detector beeps\\
$B$ :&  The ``right particle has spin `up' along direction ${\bf b} $" 
detector beeps\\
$B'$ :& The ``right particle has spin `up' along direction ${\bf b'} $" 
detector beeps\\
$a$ :& The left switch selects direction ${\bf a} $ \\
$a'$ :& The left switch selects direction ${\bf a'} $ \\
$b$ :& The right switch selects direction ${\bf b} $ \\
$b'$ :& The right switch selects direction ${\bf b'} $ 
\end{tabular}
\end{center} 

For the probabilities of these events, in case of $\angle \left( {\bf 
a},{\bf a'} \right)= \angle \left( {\bf a'},{\bf b} \right)= \angle \left( {\bf 
a},{\bf b'} \right)=120^{\circ } $ and $\angle \left( {\bf b},{\bf a'} 
\right)=0$, we have
\begin{eqnarray}
\label{Facts1} 
p(A)=p(A')=p(B)=p(B') &=&\frac{1}{4},  \nonumber \\
p(a)=p(a')=p(b)=p(b') &=&\frac{1}{2},  \nonumber \\
p(A\wedge a)=p(A) &=&\frac{1}{4},  \nonumber \\
p(A'\wedge a')=p(A') &=&\frac{1}{4}, \nonumber \\
p(B\wedge b)=p(B) &=&\frac{1}{4},  \nonumber \\
p(B'\wedge b')=p(B') &=&\frac{1}{4},  \nonumber \\
p(A\wedge a')=p(A'\wedge a)=p(B\wedge b')=p(B'\wedge b) &=&0,  
\\
p(A\wedge B)= p(A\wedge B')= p(A'\wedge B')&=&\frac{3}{32},  
\nonumber \\
p(A'\wedge B)&=&0,  \nonumber \\
p(a\wedge a)= p(b\wedge b')&=&0,   \nonumber \\
p(a\wedge b)= p(a\wedge b')= p(a'\wedge b)= p(a'\wedge 
b')&=&\frac{1}{4},  \nonumber \\
p(A\wedge b)= p(A\wedge b')= p(A'\wedge b)= p(A'\wedge b') 
\nonumber \\ =p(B\wedge a)= p(B\wedge a')= p(B'\wedge a)= 
p(B'\wedge a')&=&\frac{1}{8}.  \nonumber 
\end{eqnarray}
These statistical data {\em agree\/} with quantum mechanical results, 
in the following sense: 
\begin{eqnarray}
\label{qprob} 
\frac{p(A\wedge a)}{p(a)}=tr(\hat{W}A)=\frac{p(A'\wedge 
a')}{p(a')}= tr(\hat{W}A')\nonumber \\=\frac{p(B\wedge 
b)}{p(b)}=
tr(\hat{W}B)=\frac{p(B'\wedge 
b')}{p(b')}=tr(\hat{W}B')&=&\frac{1}{2},  \nonumber \\
\frac{p(A\wedge B\wedge a\wedge b)}{p(a\wedge b)}= 
\frac{p(A\wedge B)}{p(a\wedge b)}=tr(\hat{W}AB)  
\nonumber \\
=\frac{1}{2}\sin^{2}\angle ({\bf a},{\bf b})&=&\frac{3}{8},  
\nonumber \\  
\frac{p(A\wedge B'\wedge a\wedge b')}{p(a\wedge b')}= 
\frac{p(A\wedge B')}{p(a\wedge b')}=tr(\hat{W}AB')  
\nonumber \\
=\frac{1}{2}\sin^{2}\angle ({\bf a},{\bf b'})&=&\frac{3}{8},\\
\frac{p(A'\wedge B\wedge a'\wedge b)}{p(a'\wedge b)}= 
\frac{p(A'\wedge B)}{p(a'\wedge b)}=tr(\hat{W}A'B) 
\nonumber \\
=\frac{1}{2}\sin^{2}\angle ({\bf a'},{\bf b})&=&0,  \nonumber \\
\frac{p(A'\wedge B'\wedge a'\wedge b')}{p(a'\wedge b')}= 
\frac{p(A'\wedge B')}{p(a'\wedge b')}=tr(\hat{W}A'B')  
\nonumber \\
=\frac{1}{2}\sin^{2}\angle ({\bf a'},{\bf b'})&=&\frac{3}{8},  
\nonumber 
\end{eqnarray}
where the outcomes are identified with the following projectors
\begin{eqnarray*}
A&=&\hat{P}_{span\big\{ \psi _{+{\bf a}}\otimes \psi _{+{\bf 
a}},  \psi _{+{\bf a}}\otimes \psi _{-{\bf a}  }\big\} },\\
A'&=&\hat{P}_{span\big\{ \psi _{+{\bf a'}}\otimes \psi _{+{\bf 
a'}},  \psi _{+{\bf a'}}\otimes \psi _{-{\bf a'}  }\big\} },\\
B&=&\hat{P}_{span\big\{ \psi _{-{\bf b}}\otimes \psi _{+{\bf 
b}},  \psi _{+{\bf b}}\otimes \psi _{+{\bf b}  }\big\} },\\
B'&=&\hat{P}_{span\big\{ \psi _{-{\bf b'}}\otimes \psi _{+{\bf 
b'}},  \psi _{+{\bf b'}}\otimes \psi _{+{\bf b'}  }\big\} }
\end{eqnarray*}
of the Hilbert space $H^{2}\otimes H^{2}$. The state of the system is 
assumed to be represented by $\hat{W}=\hat{P}_{\Psi _{s}}$, where 
$\Psi _{s}=\frac{1}{\sqrt{2}}\left( \psi _{+{\bf a}}\otimes \psi _{-{\bf 
a} }-\psi _{-{\bf a}}\otimes \psi _{+{\bf a} }\right) $.

The question we would like to answer is  whether the above 
probabilities, measured in the Aspect experiment, can be 
accommodated in a Kolmogorovian probability model, or not.

\subsection{The Pitowsky formalism} 
\label{Pitowsky} 

Pitowsky  elaborated a convenient geometric language 
for the discussion of the problem whether empirically given 
probabilities are Kolmogorovian or not (Pitowsky, 1989).

Let $S$ be a set of pairs of integers $ S\subseteq  \left\{ \left\{i,j\right\} 
\mid  1 \leq i < j \leq n\right\}$. Denote by $R(n,S)$ the linear space of 
real vectors having a form like $ (f_1,f_2,...f_{ij},...)$. For each 
$\varepsilon \in \{0,1\}^n$, let $u^{\varepsilon }$  be the following 
vector 
in $R(n,S)$:

\begin{eqnarray*}
\mbox{
$\begin{array}{rclc}
 u^{\varepsilon}&=&\varepsilon_i , \hskip 1.5cm &1\leq i\leq n ,\\
 u^{\varepsilon}_{ij}&=&\varepsilon_i\varepsilon_j , &\{i,j\}\in S .
\end{array}$
}
\end{eqnarray*}

\begin{D}
\label{ D-C(n)}
The {\em classical correlation polytope\/} ${\cal C}(n,S)$ is the closed 
convex hull in  $R(n,S)$ of vectors 
$\left\{u^{\varepsilon}\right\}_{\varepsilon \in 
\{0,1\}^n}$ : 
\begin{eqnarray*}
&{\cal C}(n,S):=
&\left\{ a\in R(n,S)\mid  a=\sum_{\varepsilon \in 
\{0,1\}^n} 
\lambda_{\varepsilon}u^{\varepsilon}, \ where\ \lambda_{\varepsilon} 
\geq 0\ 
and\ 
\sum_{\varepsilon \in\{0,1\}^n} \lambda_{\varepsilon}=1 \right\}.
\end{eqnarray*}

\end{D}
Consider now events $A_{1}, A_{2}, \ldots A_{n}$  and  some of 
their conjunctions $A_{i}\wedge A_{j}\: \left( \{i,j\}\in S\right) $. 
Assume that we know their probabilities from which we can form a so 
called {\em correlation vector\/}:
\begin{eqnarray*}
{\bf p}&=&\left( p_{1}, p_{2},\ldots p_{n}, \ldots p_{ij}, \ldots \right) 
\\&=&\left(p(A_{1}), p(A_{2}), \ldots, p(A_{n}), \ldots 
p(A_{i}\wedge 
A_{j}), \ldots \right)\in R(n,S).
\end{eqnarray*}

\begin{D}
\label{KolmRep}
We will then say that ${\bf p} $ {\em has a Kolmogorovian 
representation\/} if there exist a Kolmogorovian probability space 
$(\Omega, \Sigma, \mu)$ and measurable subsets 
$$X_{A_{1}}, X_{A_{2}}, \ldots X_{A_{n}}\in \Sigma ,$$ 
such that 
\begin{eqnarray*}
\mbox{$\begin{array}{rcll}
p_{i}&=&\mu \left(X_{A_{i}}\right), \hskip 1.5cm&1\leq i\leq n,\\
p_{ij}&=&\mu \left(X_{A_{i}}\cap X_{A_{j}}\right), &
\{i,j\}\in S.
\end{array}$}
\end{eqnarray*}
\end{D}
Pitowsky's theorem tells us the necessary and sufficient condition a 
correlation vector must satisfy in order to be Kolmogorovian. 
\begin{T}[Pitowsky, 1989]
\label{PitowskyTheorem}
A correlation vector $${\bf p}=\left( p_{1}, p_{2},\ldots p_{n}, \ldots 
p_{ij}, \ldots \right)$$ has a Kolmogorovian representation if and only 
if 
${\bf p} \in {\cal C}(n,S)$.
\end{T}

In case $n=4$  and $S=S_{4}= \big\{1,3\}, \{1,4\}, \{2,3\}, 
\{2,4\}\big\} $ the condition ${\bf p} \in {\cal C}(n,S)$ is equivalent 
with the following inequalities: 
\begin{eqnarray}
\label{Eq-Clauser_Horne} 
& &0\leq p_{ij}\leq p_{i}\leq 1,  \nonumber \\
& &0\leq p_{ij}\leq p_{j}\leq 1, \hskip 1cm i=1,2\; \;j=3,4  \nonumber 
\\
& &p_{i}+p_{j}-p_{ij}\leq 1,  \nonumber \\
& &-1\leq p_{13}+ p_{14}+ p_{24}- p_{23}- p_{1}- p_{4}\leq 0,\\
& &-1\leq p_{23}+ p_{24}+ p_{14}- p_{13}- p_{2}- p_{4}\leq 0,  
\nonumber \\
& &-1\leq p_{14}+ p_{13}+ p_{23}- p_{24}- p_{1}- p_{3}\leq 0,  
\nonumber \\
& &-1\leq p_{24}+ p_{23}+ p_{13}- p_{14}- p_{2}- p_{3}\leq 0.  
\nonumber 
\end{eqnarray}
(\ref{Eq-Clauser_Horne}) reminds us the well known 
{\em Clauser-Horne  inequalities\/} (Clauser and Shimony 
1978).\footnote{There is, however,  an important conceptual 
disagreement between (\ref{Eq-Clauser_Horne}) and the original 
Clauser-Horne inequalities, see Szab\'o (1995b).}

Let us now apply inequalities (\ref{Eq-Clauser_Horne}) to the Aspect 
experiment in the usual 
way: Let
$$p_{1}=tr(\hat{W} A), p_{2}=tr(\hat{W} A'), p_{3}=tr(\hat{W} B), 
p_{4}=tr(\hat{W} B'),$$ 
$$ p_{13}=tr(\hat{W} A\hat{ B}), p_{14}=tr(\hat{W} A B'), 
p_{23}=tr(\hat{W} A'B), p_{24}=tr(\hat{W} A'B'),$$ 
the values of which are given in (\ref{qprob}). 
Substituting these values into the last inequality of 
(\ref{Eq-Clauser_Horne}) we find that
\begin{eqnarray}
\label{notin} 
{\bf p}=\left(\frac{1}{2}, \frac{1}{2}, \frac{1}{2}, 
\frac{1}{2},\frac{3}{8}, \frac{3}{8}, 0, \frac{3}{8}\right) \not\in {\cal 
C}(n,S).
\end{eqnarray}
Thus, we can draw the  {\em usual} conclusion: {\em the probabilities 
observed in the Aspect experiment have no Kolmogorovian 
representation\/}. 

However, as I pointed out in my (1995a,b), a closer analysis yields a
different conclusion! On the basis of particular examples I formulated 
the following hypothesis (Cf. 1995a). There is a ``Kolmogorovian 
Censorship" in the real world: We never encounter ``naked" quantum 
probabilities in reality. A correlation vector consisting of empirically 
testable probabilities is always a product
\begin{eqnarray*}
 \left( p_{1}\ldots p_{n}\ldots p_{ij}\ldots \right)&=& \left( \pi 
_{1}\ldots \pi _{n}\ldots \pi _{ij}\ldots \right)\cdot \left( 
\tilde{p}_{1}\ldots \tilde{p}_{n}\ldots \tilde{p}_{ij}\ldots 
\right)\\&=& \left( \pi _{1}p_{1}\ldots \pi _{n}p_{n}\ldots \pi 
_{ij}p_{ij}\ldots \right),
\end{eqnarray*}
where  $( \pi _{1}\ldots \pi _{n}\ldots \pi _{ij}\ldots) $ are quantum 
probabilities and $(\tilde{p}_{1}\ldots \tilde{p}_{n}$ $\ldots  
\tilde{p}_{ij}\ldots)$  are classical probabilities with which the 
corresponding measurements happen to be performed. The hypothesis 
says that such a product is always classical. (From the pure 
mathematical point of view, a product of a quantum and a classical 
correlation vector is not necessarily classical.) 
One can prove such a theorem within the framework of quite general 
assumptions (See Bana and Durt 1996).

Again, we must notice that Pitowsky theorem and, consequently, 
inequalities (\ref{Eq-Clauser_Horne}) apply to probabilities understood 
under  {\em one common\/} system of conditions. Thus, we make a 
serious mistake by substituting conditional probabilities (\ref{qprob}) 
into inequalities (\ref{Eq-Clauser_Horne}), since these conditional 
probabilities belong to {\em different\/} conditions.

\section{How to join probability models?}
\label{Howtojoin}
\subsection{How to do it in the classical theory?}
As we can see in the above examples, quantum mechanics produces 
Kolmogorovian probabilities belonging to different sets of conditions. 
The alleged impossibility to put these classical probability measures  
together into {\em one common\/} Kolmogorovian probability model is 
what urges us to cry for Quantum Probability Theory and Quantum 
Logic. But we {\em can\/} join these probability measures, if we do it 
in a correct way!

Before to seeing how we can do that, let us consider how this 
procedure goes  in the classical theory of probability.

\setlength{\picw}{7cm}
\setlength{\picfr}{\picw}
\addtolength{\picfr}{0.5cm}
\begin{figure}
\begin{center}
\makebox[\picfr][c]{\vbox{\centerline{\epsfxsize=\picw 
\epsfbox{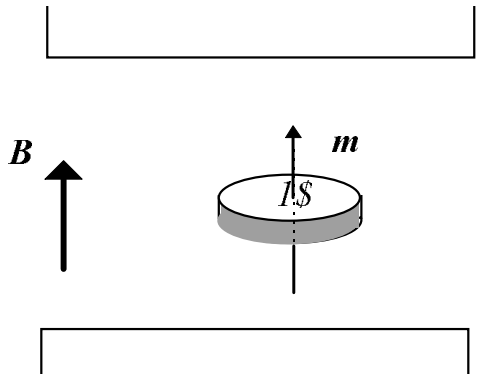}}}}
\end{center}
\caption{The probabilities of Heads (H) and Tails (T) are different if 
the magnetic field is on.}
\label{coinfig}
\end{figure}

Let me take a simple example. We are tossing a coin which has a little 
magnetic momentum (Fig.~\ref{coinfig}). If the magnetic field is off, 
the probabilities are
\begin{eqnarray*}
p_{{\rm off } }(\mbox{H}) &=& 0.5,\\
p_{{\rm off } }(\mbox{T}) &=& 0.5.
\end{eqnarray*}
If the magnetic field is on, the probabilities are different:
\begin{eqnarray*}
p_{{\rm on } }(\mbox{H}) &=& 0.2,\\
p_{{\rm on } }(\mbox{T}) &=& 0.8.
\end{eqnarray*}

\setlength{\picw}{250pt}
\setlength{\picfr}{\picw}
\addtolength{\picfr}{0.5cm}
\begin{figure}
\begin{center}
\makebox[\picfr][c]{\vbox{\centerline{\epsfxsize=\picw 
\epsfbox{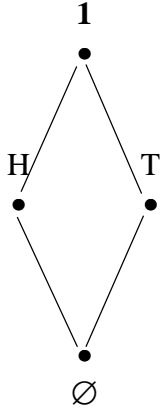}}}}
\end{center}
\caption{Algebra of events ${\cal A}$}
\label{kicsihalo}
\end{figure}

The event algebra  ${\cal A} $ is shown in Figure~\ref{kicsihalo}. 
Probability models $({\cal A}, p_{{\rm off } })$ and $({\cal A}, 
p_{{\rm on } })$ are, separately, Kolmogorovian. For example, they 
satisfy inequality~(\ref{1}):
\begin{eqnarray}
\label{1off}
p_{{\rm off }}({\rm H }) + p_{{\rm off } }({\rm T })-p_{{\rm off 
}}({\rm H } \wedge {\rm T }) \leq 1,
\end{eqnarray}
and separately,
\begin{eqnarray}
\label{1on}
p_{{\rm on } }(H)+p_{{\rm on}}(T)-p_{{\rm on } }(H\wedge T) \leq 
1.
\end{eqnarray}
If we make the same mistake we did in the previous  examples,  
and put these probabilities, belonging to different conditions, together 
into one formula prescribed for a Kolmogorovian probability theory, 
we find the same kind of  ``violation of the rules of classical 
probability theory":
\begin{eqnarray}
\label{absurd}
p_{{\rm off } }(\mbox{ H })+p_{{\rm on } }({\rm T }))-p_{{\rm off } 
}({\rm H } \wedge {\rm T }) = 0.5 + 0.8 > 1, 
\end{eqnarray}
or
\begin{eqnarray}
\label{absurd2}
p_{{\rm on } }({\rm H }) + p_{{\rm off } }({\rm T }) = 0.2 + 0.5 \neq 
1 = p_{{\rm off } }(1) = p_{{\rm off } }( {\rm H }\vee {\rm T }).
\end{eqnarray}

\setlength{\picw}{200pt}
\setlength{\picfr}{\picw}
\addtolength{\picfr}{0.5cm}
\begin{figure}
\begin{center}
\makebox[\picfr][c]{\vbox{\centerline{\epsfxsize=\picw 
\epsfbox{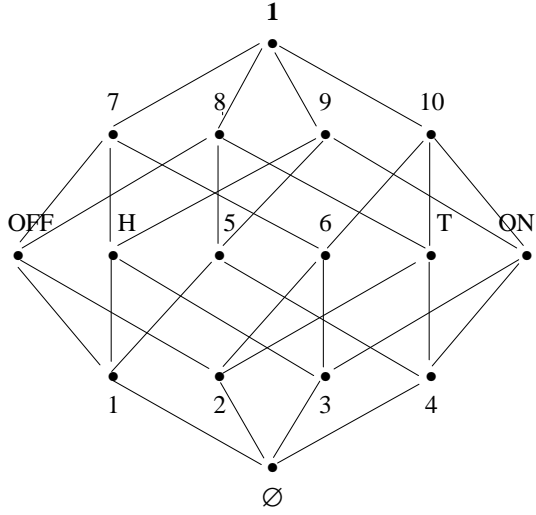}}}}
\end{center}
\caption{The unified algebra of events ${\cal A'}$}
\label{nagyhalo}
\end{figure}

Consider now how to join probability models $({\cal A}, p_{{\rm off } 
})$ and $({\cal A}, p_{{\rm on } })$. In the classical probability theory 
we can join probabilities belonging to separate conditions only by 
enlarging the event algebra in such a way that it contains not only the 
original events but the ``conditioning events", too 
(Fig.~\ref{nagyhalo})\footnote{I am grateful to Miltos Zissis for his warning 
that Fig.~\ref{nagyhalo} was incorrect in a previous version of this paper}. 
Of course, we can do that only if we know the 
probabilities of the conditioning events. In the example of question 
assume that $p(\mbox{OFF})=0.5$ and $p(\mbox{ON})=0.5$. So, the 
unified probability model is $({\cal A'}, p)$, where
\begin{eqnarray}
\label{uniprob}
p({\rm 1})=p({\rm 2})= p({\rm 9})= p({\rm 10})&=&0.25,  \nonumber \\
p({\rm 3})= p({\rm 8})&=&0.1,  \nonumber \\
p({\rm 4})= p({\rm 7})&=&0.4,  \nonumber \\
p({\rm OFF })=p({\rm ON })&=&0.5,   \\
p({\rm H })= p({\rm 6})&=&0.35,   \nonumber \\
p({\rm T })= p({\rm 5})&=&0.65.  \nonumber 
\end{eqnarray}
The original probabilities are represented as conditional probabilities 
(defined by the Bayes law):
\begin{eqnarray}
\label{condrep}
p_{{\rm on } }({\rm H })&=&\frac{p({\rm H }\wedge {\rm ON 
})}{p({\rm ON })}=\frac{p({\rm 3})}{p({\rm ON 
})}=\frac{0.1}{0.5}=0.2,  \nonumber \\
p_{{\rm on } }({\rm T })&=&\frac{p({\rm T }\wedge {\rm ON 
})}{p({\rm ON })}=\frac{p({\rm 4})}{p({\rm ON 
})}=\frac{0.4}{0.5}=0.8,  \nonumber \\
p_{{\rm off } }({\rm H })&=&\frac{p({\rm H }\wedge {\rm OFF 
})}{p({\rm OFF })}=\frac{p({\rm 1})}{p({\rm OFF 
})}=\frac{0.25}{0.5}=0.5, \\
p_{{\rm off } }({\rm T })&=&\frac{p({\rm T}\wedge {\rm OFF 
})}{p({\rm OFF })}=\frac{p({\rm 2})}{p({\rm OFF 
})}=\frac{0.25}{0.5}=0.5.  \nonumber 
\end{eqnarray}

\subsection{How to do it in the quantum theory?}

Consider a quantum system described in Hilbert space $H$. The state 
of the system is represented by density operator $W$. Assume that 
there are  $N$ different measurements $m_{1}, m_{2}, \ldots m_{N}$ 
one can carry out on the system. The corresponding 
observable-operators are denoted by $\hat{M}_{1}, \hat{M}_{2}, 
\ldots \hat{M}_{N}$.  Let ${\cal M}_{1}, {\cal M}_{2}, \ldots {\cal 
M}_{N}$ be the  spectra of these operators. Introduce the following 
notation: each set of measurements $\big\{m_{i_{1}},\ldots 
m_{i_{s}}\big\}$ will be identified with a vector $\eta \in 
\{0,1\}^{N}$, such that 
\begin{eqnarray*}
\eta _{i}= \left\{\: \begin{array}{lll}
1&\mbox{if} &m_{i}\in \big\{m_{i_{1}},\ldots m_{i_{s}}\big\},\\
0&\mbox{if} &m_{i}\not\in \big\{m_{i_{1}},\ldots m_{i_{s}}\big\}.
\end{array}\right.
\end{eqnarray*}
In this way the conditioning events can be represented in 
$2^{\{0,1\}^{N}}$. For instance, event  ``measurement $m_{i}$ is 
performed" is represented by $\big\{\eta \left|  \eta _{i}=1\right. 
\big\}\subset \{0,1\}^{N}$, event ``measurement $m_{i}$ and 
measurement $m_{j}$ are performed" corresponds to $$\big\{\eta \left|  
\eta _{i}=1\right.\big\} \cap \big\{\eta \left|  \eta _{j}=1\right.\big\},$$ 
etc.

 Some of these measurements can be incompatible, in the sense that 
they cannot be simultaneously carried out. Assume that, according to 
the quantum theory, observables belonging to compatible 
measurements commute.
For each set $\{m_{i_{1}}, m_{i_{2}}, \ldots m_{i_{s}}\}= 
\{m_{i}\}_{\eta _{i}=1} $ of  {\em compatible\/} measurements the 
quantum state $W$ determines a Kolmogorovian probability measure 
over the corresponding Borel sets, $\Big(B\Big(\begin{array}[t]{c} 
\mbox{{\Huge $\times$}}\\^{\eta _{i}=1}\end{array}{\cal 
M}_{i}\Big),\, \mu_{\eta}\Big)$, where
\begin{eqnarray}
\label{quantkolmmeasure}
\mu_{\eta}:  (A_{i})_{\eta _{i}=1}\in B\Big(\begin{array}[t]{c} 
\mbox{{\Huge \boldmath $\times$}}\\^{\eta _{i}=1}\end{array}{\cal 
M}_{i}\Big) \mapsto tr\left(W \prod_{\eta _{i}=1}A_{i}\right).
\end{eqnarray}

Now, how can we join these classical probability spaces into one 
common classical probability model? The method is known from  the 
classical theory of probability. Quantum mechanics has nothing special 
from this point of view! That is, we need to enlarge the event algebra 
by the conditioning events and to define the joint probability measure 
over this larger algebra of events. In order to do that, we need to know 
the probabilities of  conditioning events. The values of these 
probabilities are the matter of empirical facts. Although, the following 
assumption seems to be quite plausible:
\begin{ST}
\label{empiria}
There is a classical probability measure $\tilde{p}$ on  
$2^{\{0,1\}^{N}}$, such that
if $\tilde{p}\left(\left\{\eta \right\} \right) \neq 0$ then the 
corresponding set of operators $\big\{\hat{M}_{i}\big\}_{ \eta 
_{i}=1}$ is commuting.
\end{ST}

Thus, my assertion is that classical probabilities 
(\ref{quantkolmmeasure}) can be joint into one Kolmogorovian 
probability model:

\begin{T}
\label{unification}
There exists a Kolmogorovian probability space 
$\big({\cal M}, B({\cal M}), p\big)$ such that each conditioning event 
$E \in 2^{\{0,1\}^{N}}$ and each outcome event 
$(A_{i})_{\eta_{i}=1}$ can be represented by an element of 
$ B({\cal M})$, denoted by $X_{E}$ and 
$X_{(A_{i})_{\eta_{i}=1}}$, respectively, and 
\begin{eqnarray*}
\begin{array}{rclcl}
\tilde{p}(E) &=& p(X_{E}),
&\mbox{\hskip1cm}&\  \forall E\in 2^{\{0,1\}^{N}},\\
\ \\
\ \\
\mu_{\eta} \big((A_{i})_{\eta _{i}=1}\big) &=& tr\Big( W\prod_{\eta 
_{i}=1}A_{i}\Big) \\&=&
\frac{ p\left( X_{(A_{i})_{\eta _{i}=1}}\cap X_{\{\eta 
\}}\right)}{p\left( X_{\{\eta \}}\right)},
& &
\begin{array}{l}
\forall \eta \in \{0,1\}^{N} \\
\forall (A_{i})_{\eta _{i}=1}\in \begin{array}[t]{c}\mbox{{\Huge 
$\times$}}\\^{\eta_{i}=1}\end{array}{\cal M}_{i}.
\end{array}
\end{array}
\end{eqnarray*}
\end{T}

\begin{Proof}
\normalsize 
The enlarged Boolean algebra of events can be constructed as a 
Boolean $\sigma $-algebra of Borel sets $B({\cal M})$, where 
\begin{eqnarray}
\label{BigSet}
{\cal M}= \bigcup_{\eta \in 
\{0,1\}^{N}}^{disjoint}\begin{array}[t]{c}\mbox{{\Huge 
$\times$}}\\^{\eta_{i}=1}\end{array} {\cal M}_{i}.
\end{eqnarray}
 
An original outcome event $A_{i}\in B({\cal M}_{i})$ is represented 
by
$$\bigcup^{disjoint}_{\begin{array}{c}
\scriptstyle{\eta \in \{0,1\}_{N}}\\\scriptstyle{\eta _{i}=1}
\end{array}}\Big(\begin{array}[t]{c}\mbox{{\Huge 
$\times$}}\\^{\begin{array}{c}
\scriptstyle{\eta _{j}=1}\\\scriptstyle{ j<i }
\end{array} }\end{array} {\cal M}_{j}\Big) \  \times \  A_{i} \  \times 
\  \Big(\begin{array}[t]{c}\mbox{{\Huge 
$\times$}}\\^{\begin{array}{c}
\scriptstyle{\eta _{k}=1}\\\scriptstyle{ k>i }
\end{array} }\end{array} {\cal M}_{k}\Big). $$ 

A conditioning event $\left\{ \eta \right\}  $ can be identified with 

$$\begin{array}[t]{c}\mbox{{\Huge $\times$}}\\^{\eta 
_{i}=1}\end{array} {\cal M}_{i}.$$

Now, the joint classical probability model is $\big({\cal M}, B({\cal 
M}), p \big)$, where the probability measure  $p $ is generated by the 
following rule:

\begin{eqnarray}
\label{jointprob}
p:\left(\begin{array}[t]{c}\mbox{{\Huge $\times$}}\\^{\eta 
_{i}=1}\end{array} A_{i}\right) \in B({\cal M}) \mapsto tr\left(W  
\prod_{\eta _{i}=1}A_{i}\right)\cdot \tilde{p}(\eta ). 
\end{eqnarray}
It is worth mentioning that if  $\eta $ represents incompatible 
measurements then probability (\ref{jointprob}) is zero.

The original probabilities belonging to different particular condition 
are also reproduced as conditional probabilities: for example the 
probability of an outcome $A_{i}$ given that measurement $m_{i}$ is 
carried out is $tr(WA_{i})$, and indeed,

\begin{eqnarray}
\label{reproducedprob}
& &\frac{ p\Big(\bigcup^{disjoint}_{ \begin{array}{c}
\scriptstyle{\eta \in \{0,1\}^{N}}\\\scriptstyle{\eta _{i}=1}
\end{array}}\Big(\mbox{{\huge $\times$}}_{ \begin{array}{c}
\scriptstyle{\eta _{j}=1}\\\scriptstyle{ j<i }
\end{array} } {\cal M}_{j}\Big) \times A_{i} \  \times \  
\Big(\mbox{{\huge $\times$}}_{ \begin{array}{c}
\scriptstyle{\eta _{k}=1}\\\scriptstyle{ k>i }
\end{array} } {\cal M}_{k}\Big) \Big)}{ p\Big(\mbox{{\huge 
$\times$}}_{\eta _{i}=1} {\cal M}_{i}\Big)}  \nonumber \\
& =& \frac{\sum_{ \begin{array}{c}
\scriptstyle{\eta \in \{0,1\}^{N}}\\\scriptstyle{\eta _{i}=1}
\end{array}}tr\Big( W\Big(\prod_{\eta _{j}=1}Id_{j}\Big)A_{i} \Big) 
\tilde{p}(\eta )}{\sum_{ \begin{array}{c}
\scriptstyle{\eta \in \{0,1\}^{N}}\\\scriptstyle{\eta _{i}=1}
\end{array}}\tilde{p}(\eta )}\\
&=& \frac{tr\Big( WA_{i}\Big) \sum_{ \begin{array}{c}
\scriptstyle{\eta \in \{0,1\}^{N}}\\\scriptstyle{\eta _{i}=1}
\end{array}}\tilde{p}(\eta ) }{\sum_{ \begin{array}{c}
\scriptstyle{\eta \in \{0,1\}^{N}}\\\scriptstyle{\eta _{i}=1}
\end{array}}\tilde{p}(\eta )}= tr\left(W A_{i}\right).  \nonumber 
\end{eqnarray}

\end{Proof}
	
Let us apply this general method  to the EPR experiment. We have $4$ 
possible measurements $a, a', b, b'$.  There are 8 different conditioning 
events symbolized with the  8 vectors $\eta \in \{0,1\}^{4}$. The 
probabilities of these conditioning events are
\begin{eqnarray*}
\tilde{p}(0000)= \tilde{p}(1000)= \tilde{p}(0100)= \tilde{p}(0010)= 
\tilde{p}(0001)= \tilde{p}(1100)= \tilde{p}(0011)&=&0,\\
\tilde{p}(1010)= \tilde{p}(1001)= \tilde{p}(0110)= 
\tilde{p}(0101)&=&\frac{1}{4}.
\end{eqnarray*}

Each of the four observables has a spectrum consisting from two points 
``up" and ``down". Thus, the enlarged event algebra is the Boolean 
algebra generated by 81 elementary events $E\in \{up, down, 
none\}^{4}$.  The joint probability measure is determined by the 
probabilities of the elementary events. There are 16 elementary events 
which have non-zero probability:
\begin{eqnarray*}
&  &p(\mbox{up none up none})= p(\mbox{up none none up})\\&=& 
p(\mbox{none up none up})= p(\mbox{down none down none})\\&=&
p(\mbox{down none none down})= p(\mbox{none down none 
down})=\frac{3}{32},\\
& &p(\mbox{up none down none})= p(\mbox{up none none 
down})\\&=& p(\mbox{none up none down})= p(\mbox{down none up 
none})\\&=& p(\mbox{down none none up})= p(\mbox{none down 
none up})=\frac{1}{32},\\
& &p(\mbox{none down up none })= p(\mbox{none up down 
none})=\frac{1}{8}.
\end{eqnarray*}

To see that this is a consistent representation, let us check one of the 
probabilities in (\ref{qprob}): event $A\wedge B$ is represented by 
subset
\begin{eqnarray*}
& &X_{A}\cap X_{B}=
 \Big\{\mbox{(up none up none), (up up up none)},\\& &\mbox{(up 
down up none), (up none up down), (up none up up)}\Big\}.
\end{eqnarray*} 
The probability $p(X_{A}\cap X_{B})=\frac{3}{32} + 0 + 0 + 0 + 0=  
\frac{3}{32}$. Condition event $a\wedge b$ is represented by
\begin{eqnarray*}
& &X_{a}\cap X_{b}=\Big\{ \mbox{(up none up none), (down none 
up none)},\\ & &\mbox{(down none down none), (up none down 
none)},\\
& &\mbox{(up up up up), (up up up none), \ldots (down down down 
down)}\Big\}
\end{eqnarray*}
and $$p(X_{a}\cap X_{b})= \frac{3}{32} + \frac{1}{32} + 
\frac{3}{32} + \frac{1}{32} + 0 + 0 \ldots + 0=\frac{1}{4}.$$
As it is required, $X_{A}\cap X_{B}\subset X_{a}\cap X_{b}$.
Consequently,

\begin{eqnarray*}
tr(WAB)=\frac{p(X_{A}\cap X_{B}\cap X_{a}\cap 
X_{b})}{p(X_{a}\cap X_{b})}=\frac{p(X_{A}\cap 
X_{B})}{p(X_{a}\cap 
X_{b})}=\frac{\frac{3}{32}}{\frac{1}{4}}=\frac{3}{8},
\end{eqnarray*}
as it was expected.

\section*{Conclusions}
The analysis of the above examples and Theorem~\ref{unification} 
unanimously tell us  that quantum mechanics, regarded as a physical 
theory about empirical facts of our world, does not demand to 
supersede the classical theory of probability.
It is needless to do that and, what is more, any attempt at the empirical 
foundation of the quantum probability theory seems to be 
contradictory. Like it or not, quantum mechanics is connected with the 
empirical facts about the world, which it is supposed to be applied to, 
through relative frequencies. But those  ``probabilities" that are 
presented by the quantum probability theory can hardly be interpreted 
as relative frequencies of events. And whether we like it or not, 
quantum logic is nothing else but an algebraic structure isomorphic 
with the algebra of events underlying the quantum probability theory. 
So, if quantum probability theory has nothing to do to reality then 
quantum logic is meaningless, too. Quantum logic and quantum 
probability theory remain fascinating mathematical theories but 
without any relevance to our real world.

\section*{References}
\begin{list}%
{ }{\setlength{\itemindent}{-15pt}
\setlength{\leftmargin}{15pt}}

\item Ballentine, L. E., (1989): ``Probability theory in quantum 
mechanics", in {\em The Concept of Probability\/}, eds. E. I. Bitsakis 
and C. A. Nicolaides, Kluwer, Dordrecht.
\item Bana, G. and T. Durt, (1996): ``Proof of Kolmogorovian 
Censorship", in preparation.
\item Bell, J. S., (1987):  {\em Speakable and unspeakable in quantum 
mechanics\/}, Cambridge University Press, Cambridge 
\item Beltrametti, E. G. and G. Cassinelli, (1981): {\em The logic of 
quantum mechanics\/}, Addison-Wesley, Reading, Massachusetts.
\item Birkhoff, G. and J. von Neumann, (1936): ``The logic of quantum 
mechanics", {\em Ann. Math.\/} {\bf 37}, 823-843.
\item Bub, J., (1974): {\em The interpretation of quantum 
mechanics\/}, Reidel, Dordrecht, Holland.
\item Clauser, J. F. and  A. Shimony, (1978): ``Bell's theorem: 
experimental tests and implications", {\em Rep. Prog. Phys.\/} {\bf 
41}, 1881-1927.
\item Costantini, D., (1992): ``A statistical analysis of the two-slits 
experiment, or some remarks on quantum probability", {\em Universita 
Degli Studi Di Genova, Istituto Di Statistica\/}, preprint.
\item Einstein, A., (1938): ``Physik und Realit\"at", {\em Journ. 
Franklin Institut\/} {\bf 221}, 313-347.
\item Gill, R. D. (forthcoming): ``Critique of `Elements of Quantum 
Probability'", in {\em Quantum Probability Communications XI.\/}, ed. 
R. L. Hudson and J. M. Lindsay, World Scientific, Singapour.
\item Gleason, A. M., (1957): ``Measures on the closed subspaces of a 
Hilbert space", {\em J. math. Phys\/} {\bf 6}, 8855-893.
\item Gudder, S. P., (1988): {\em Quantum probability\/}, Academic 
Press, Boston.
\item Jauch, J. M., (1968):{\em Foundations of Quantum Mechanics\/}, 
Addison-Wesley, Reading, Massachusetts.
\item K\"ummerer, B. and  H. Maassen, (1996): ``Elements of quantum 
probability", in {\em Quantum Probability Communications X.\/}, ed. 
R. L. Hudson and J. M. Lindsay, World Scientific, Singapour.
\item Mittelsteadt, P., (1978): {\em Quantum logic\/}, Reidel, 
Dordrecht, Holland.
\item Piron, C., (1976): {\em Foundations of quantum physics\/}, 
Benjamin, Reading, Massachusestts.
\item Pitowsky, I., (1989):  {\em Quantum Probability - Quantum 
Logic\/}, Lecture Notes in Physics {\bf 321}, Springer, Berlin
\item Putnam H., (1974): ``How to think quantum-logically?", {\em 
Synthese\/}, {\bf 29}, 55-61.
\item R\'edei, M. (forthcoming): ``Why John von Neumann did not like 
the Hilbert space formalism of quantum mechanics (and what he liked 
instead)",  {\em Studies in the History and Philosophy of Modern 
Physics\/}
\item Strauss, M., (1937): ``Matematics as logical syntax --- A method 
to formalize the language of a physical theory, {\em Erkenntnis\/}, {\bf 
7}, 147-153.
\item Szab\'o, L. E., (1995a): ``Is quantum mechanics compatible with  
a deterministic universe? Two interpretations of quantum  
probabilities" {\it Foundations of Physics Letters}, {\bf 8}, 421-440.
 \item Szab\'o, L. E., (1995b): ``Quantum mechanics in an entirely 
deterministic universe" {\it Int. J. Theor. Phys.}, {\bf 34}, 1751-1766.
\item von Neumann, J., (1932): {\em Mathematische Grundlagen der 
Quantenmechanik\/}, Springer, Berlin.

\end{list}

\end{document}